\documentclass{mem}
\usepackage{natbib}\usepackage{txfonts}\usepackage{balance}
\usepackage{graphicx}
\usepackage[a4paper]{hyperref}
\idline{1}{1}
\begin{document}

\title{Varying Constants: Constraints from Seasonal Variations}
\author{
D.\, J.\,Shaw\inst{1} 
\and J.\, D.\, Barrow\inst{2}
          }

  \offprints{D. J. Shaw}
\institute{
School of Mathematical Sciences, 
Queen Mary University of London, 
London E1 4NS, UK
\email{D.Shaw@qmul.ac.uk}
\and
DAMTP, Centre for Mathematical Sciences, 
University of Cambridge, Wilberforce Road, 
Cambridge CB3 0WA, UK
}
\authorrunning{Shaw }
\titlerunning{Constraints from Seasonal Variations}
\abstract{
We analyse the constraints obtained from new atomic clock data on the
possible time variation of the fine structure `constant' and the
electron-proton mass ratio and show how they are strengthened when the
seasonal variation of Sun's gravitational field at the Earth's surface is
taken into account. }

\maketitle

General relativity and the standard model of particle physics depend on some 27 seemingly independent numerical parameters.  These include the fine structure constants with determine the strengths of the different forces, matrix angles and phases and the relative masses of all known fundamental particles.   Together, these parameters are commonly referred to as the `fundamental constants of Nature', although many modern proposals for fundamental physics predict that they are neither strictly fundamental nor constant.

Indeed, variation of the traditional `constants', at some level, is a common prediction of most modern proposals for fundamental physics beyond the Standard Model.  For instance, if the true fundamental theory exists in more than four
space-time dimensions, the constants we observe are merely four-dimensional `shadows' of the truly fundamental higher dimensional constants. The four dimensional constants will then be seen to vary as the extra dimensions change slowly in size.  Searches for, and limits on, the variations of the traditional constants therefore provide an important probe of fundamental physics.

In many theoretical models of varying constants one expects the values of the constants to evolve slowly as the Universe expands.  Locally, this would manifest itself as a slow temporal drift in the values of the constants.  Laboratory constraints on such a drift are generally found by comparing clocks based on different atomic frequency standards
over a period of one or more years. The current best limit on the drift of the electromagnetic fine structure constant, $\alpha$,  was found by \citet{Rosenband}. The ratio of aluminium and mercury single-ion optical clock frequencies, $f_{\mathrm{Al+}%
}/f_{\mathrm{Hg+}}$, was repeatedly measured over a period of about a year and these measurements gave: $\dot{\alpha}/\alpha =
(-5.3\pm 7.9)\times 10^{-17}\,\mathrm{yr}^{-1}$. 

If the `constants', such as $\alpha $, can vary, then in addition to a slow temporal drift one would also expect to see
an annual modulation in their values \citep{seasonal}. In many theories, the Sun perturbs the values of the constants by a factor roughly proportional to the Sun's Newtonian gravitational potential, which scales as the inverse of distance, $r$, between the Earth and the Sun.  Since $r$ fluctuates annually, reaching a minimum at perihelion in early January and a maximum at aphelion in July, the values of the constants, as measured here on Earth, should also
oscillate in a similar seasonal manner.  Moreover, in many models, this seasonal fluctuation is predicted to dominate  over any linear temporal drift \citep{seasonal,seasonal2}.

We suppose that the Sun creates a distance-dependent
perturbation to the measured value of a coupling constant, $\mathcal{C}$, of
amplitude $\delta \ln \mathcal{C}=C(r)$. If this coupling constant is
measured on the surface of another body (e.g. the Earth) which orbits the first
body along an elliptical path with semi-major axis $a$, period $T_{\mathrm{p}%
}$, and eccentricity $e\ll 1$, then to leading order in $e$, the annual
fluctuation in $\mathcal{C}$, $\delta \mathcal{C}_{\mathrm{annual}}$ will be
given by 
\begin{equation}
\frac{\delta \mathcal{C}_{\mathrm{annual}}}{\mathcal{C}}=-c_{\mathcal{C}%
}\cos \left( \frac{2\pi t}{T_{\mathrm{p}}}\right) +O(e^{2}),  \label{VarForm}
\end{equation}%
where $c_{\mathcal{C}}\equiv e\,a C^{\prime }(a)$, $C^{\prime }(a) = d C(r)/dr \vert_{r=a}$ and $t=nT_{\mathrm{p}}$,
for any integer $n$, corresponds to the moment of closest approach
(perihelion). For the Earth, $a=149,597,887.5\,\mathrm{km}$. In the case of the Earth moving around the Sun, over a period of 6 months from perihelion to aphelion one would therefore measure a change
in the constant $\mathcal{C}$ equal to $2c_{\mathcal{C}}$.   As we noted above, in most theoretical models of varying constants, $\delta \ln \mathcal{C} = C(r) \propto \Delta U_{\odot}(r)$ where $U_{\odot}(r)=-GM/r$ is the Newtonian potential of the Sun.  We introduce sensitivity parameters $k_{\mathcal{C}}$ defined by $\delta \ln \mathcal{C} = k_{\mathcal{C}} \Delta U_{\odot}$. Hence:
\begin{equation}
c_{\mathcal{C}} =ea \frac{GM_{\odot}}{a^2}k_{\mathcal{C}} = 1.65\times 10^{-10}k_{\mathcal{C}}.
\end{equation}
Once one specifies a theoretical model of varying  constants, the $k_{\mathcal{C}}$ are determined. Hence measuring or bounding the $k_{\mathcal{C}}$ constrains the underlying theory.  One generally expects $k_{\mathcal{C}} \sim O(1)$ in higher dimensional theories such as String Theory, however, as we shall see, observations typically constrain $k_{\mathcal{C}} \ll 1$.

\begin{figure}[]
\resizebox{\hsize}{!}{\includegraphics[scale=0.40]{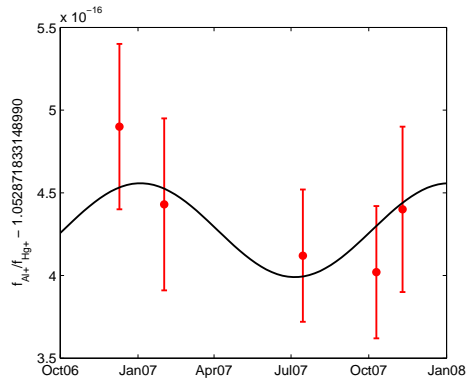}}
\caption{\footnotesize Frequency ratio $f_{\mathrm{Al+}}/f_{\mathrm{Hg+}}$
as measured by \citet{Rosenband}. The solid
black line shows the maximum likelihood fit for a seasonal variation.}
\label{fig1}
\end{figure}
\citet{Rosenband} fitted a linear drift in $\alpha $
to their data finding $\dot{\alpha}/\alpha =(-1.6\pm 2.3)\times 10^{-17}\,%
\mathrm{yr}^{-1}$. We fitted the expected form of any annual fluctuation, Eq. %
\ref{VarForm}, to the measured values of $f_{\mathrm{Al+}}/f_{\mathrm{Hg+}}$%
 \citep{seasonal2}.  Fig. \ref{fig1} shows the best-fit line superimposed on the \citep{Rosenband} data. It should be noted that the magnitudes of the systematic errors for the middle three data points were not verified to the same precision as they were from the first and last data points \citep{Rosenband}. We do not expect to this have a great effect on the resulting constraint on $k_{\alpha}$. Using $\delta \ln (f_{\mathrm{Al+}}/f_{\mathrm{Hg+}%
})=(3.19+0.008)\delta \alpha /\alpha $, \citep{Rosenband}, a maximum
likelihood fit to the data gives 
\begin{equation}
k_{\alpha }=(-5.4\pm 5.1)\times 10^{-8}.  \label{kalpha}
\end{equation}%
This limit on $k_{\alpha}$ is almost an 2 order of magnitude improvement on the previous limit of $k_{\alpha }=(2.5\pm 3.1)\times 10^{-6}$ from by \citet{blatt}.  

The frequency shifts measured by \citet{Rosenband} are not sensitive to changes in the electron-proton mass ratio: $\mu =m_{%
\mathrm{e}}/m_{\mathrm{p}}$. However, measurements of optical transition frequencies
relative to Cs are sensitive
to both $\mu $ and $\alpha $, and H-maser atomic clocks can detect changes in the light quark to proton mass ratio: $q=m_{%
\mathrm{q}}/m_{\mathrm{p}}$.   We define sensitivity parameters $k_{\mu}$ and $k_{q}$ for $\mu$ and $q$ respectively. We then find that the $\mathrm{Yb}^{+}$
frequency measurements of \citet{peik} give $k_{\alpha }+0.51k_{\mu }=(7.1\pm 3.4)\times 10^{-6}$.  Combining this limit with that on $k_{\alpha}$ and the limits $k_{\alpha }+0.36k_{\mu }=(-2.1\pm 3.2)\times 10^{-6}$ \citep{blatt},  $k_{\alpha
}+0.17k_{\mu }=(3.5\pm 6.0)\times 10^{-7}$ \citep{fortier} and  $k_{\alpha }+0.13k_{\mathrm{q%
}}=(1\pm 17)\times 10^{-7}$ \cite{Ashby}, we find \citep{seasonal2}
\begin{eqnarray}
k_{\mu } &=&(3.9\pm 3.1)\times 10^{-6},  \label{kmu} \\
k_{\mathrm{q}} &=&(0.1\pm 1.4)\times 10^{-5}.  \label{kq}
\end{eqnarray}
Again these represent improvements on the previous limits of $k_{\mu } =(-1.3\pm 1.7)\times 10^{-5}$ and $k_{\rm q} =(-1.9\pm 2.7)\times 10^{-5}$ found by \citet{blatt}.

Seasonal fluctuations occur in varying constant theories because the constants depend on the the vacuum expectation value one or more scalar fields, $\phi_I$, which interact with normal matter.  In the solar system the field equations of these scalars typically reduce to Poisson equations, similar to that satisfied by the Newtonian gravitational potential. We label the different constants $\mathcal{C}_{j}(\phi_I)$ and then:
\begin{equation}
\nabla^2 \phi_{I} \approx 4\pi G\sum_{j}  \frac{\partial \ln \mathcal{C}_{j}}{\partial \phi_{I}} \frac{\delta \rho}{\delta \ln \mathcal{C}_{j}}.\nonumber
\end{equation}
We define $\beta_{Ij} = \partial \ln \mathcal{C}_{j}/\partial \phi_{I}$. For small variations of the scalar fields and the constants we have:
\begin{equation}
\nabla^2 \ln \mathcal{C}_{i} \approx \left(\sum_{I, j} \beta_{Ii}\beta_{Ij} \frac{\delta \ln \rho}{\delta \ln \mathcal{C}_{j}}\right) 4\pi G\rho.
\end{equation}
The Newtonian potential, $\Phi_{N}$, obeys: $\nabla^2 \Phi_N = 4\pi G \rho$, and the $k_{\mathcal{C}_{i}}$ are defined by $\ln \mathcal{C}_{i} = {\rm const} + k_{\mathcal{C}_{i}}\Phi_N$. It follows that for a body orbiting the Sun with $\rho = \rho_{\odot}$:
$$
k_{\mathcal{C}_{i}} =  \left(\sum_{I, j} \beta_{Ii}\beta_{Ij} \frac{\delta \ln \rho_{\odot}}{\delta \ln \mathcal{C}_{j}}\right).
$$
We define $\lambda_{(\odot)}^{j} = \delta \ln \rho_{\odot} / \delta \ln \mathcal{C}_{j}$; these can be calculated from the standard model using information about the composition of the Sun. Measurements of $\mathcal{k}_{\mathcal{C}_{i}}$ from seasonal variations therefore determine the model parameters $\sum_{I}\beta_{Ii}\beta_{Ij}$ in a model independent fashion.

The presence of spatial gradients in the scalar fields with couple to normal matter also results in new or 	`fifth' forces.   The magnitude of the new force towards the Sun on a test body with density $\rho_{\rm 0}$, mass $m_0$ at a distance $r$ from the Sun is given by:
\begin{eqnarray}
F_{\phi} &=& m_{0} \beta_{Ij} \lambda^{(t)}_{j} \frac{\partial \phi_{I}}{\partial r} = \frac{Gm_0 M_{\odot}}{r} \sum_{j} k_{\mathcal{C}_j} \lambda^{j}_{(0)}.  
\end{eqnarray}
Now $F_{\phi}$ depends on $\lambda^{j}_{(0)} = \delta \ln \rho_{0} / \delta \ln \mathcal{C}_{j}$ which depends on the composition of the test mass; for a given test mass these can be calculated using the standard model.  This composition dependence violates the universality of free-fall and hence the weak equivalence
principle (WEP). WEP violation searches measure the differential acceleration of two differently composed test masses towards the Sun which is proportional to $\sum_{i} k_{\mathcal{C}_i}  \Delta \lambda^{i}$;  $\Delta \lambda^{i}$ is the difference between the $\lambda^{j}$ values for the two test masses. Hence WEP violation searches indirectly probe the $k_{\mathcal{C}_i}$. The magnitude of any composition-dependent fifth force
toward the Sun is currently constrained to be no stronger than about $%
{\rm few} \times 10^{-13}$ times than the gravitational force \citep{WEP}. To extract limits on the individual $k_{\mathcal{C}_i}$ one must accurately calculate the $\Delta \lambda^{i}$ parameters for different combinations of  test masses and have limits from different experiments to eliminate the degeneracy between the different $k_{\mathcal{C}_{i}}$.  

The constraints from WEP tests indirectly bound $%
k_{\alpha }$. Indeed, they often provide the tightest constraints on $%
k_{\alpha }$ \citep{seasonal,seasonal2,dent}. A recent and thorough analysis of the WEP violation constraints on $%
k_{\alpha }$ \citep{dent} found: 
$$
\vert k_{\alpha } \vert \lesssim  (0.23 - 1.4) \times 10^{-8}
$$
with a similar limit on $k_q$. The uncertainty in this limit is to due to how one models nuclear structure.  
Despite these uncertainties,  it is clear that WEP violation constraints from laboratory experiments currently provide the strongest, albeit indirect, bounds on $k_{\alpha }$ and the other $k_{\mathcal{C}_{i}}$.

We now consider the sensitivity that would be required of an atomic clock to provide tighter constraints on  $k_{\alpha}$ than those coming indirectly from limits on WEP violation.    Suppose that the ratio of two transition frequencies, $f_{\mathrm{A}}/f_{\mathrm{B}}
$, can be measured with a sensitivity $\sigma _{\mathrm{f}}$, and that $%
\delta \ln (f_{\mathrm{A}}/f_{\mathrm{B}})=S_{\alpha }\delta \alpha /\alpha $
(typically $S_{\alpha }\sim \mathcal{O}(1),$ although some transitions
exhibit a greatly increased sensitivity \citep{flamth}). The sensitivity to
changes in $\alpha $ is then given by $\sigma _{\alpha }=\sigma _{\mathrm{f}%
}/S_{\alpha }$. By simulating data sets, we found that the sensitivity to $%
k_{\alpha }$ is significantly improved if one makes $N_{\mathrm{m}}\gtrsim 12
$ measurements per year (at roughly regular intervals). With $N_{\mathrm{m}%
}\gtrsim 12$, by performing a bootstrap linear regression with $10^{5}$
re-samplings of the simulated data points, we find that the sensitivity, $%
\sigma _{k}$ to $k_{\alpha }$ is roughly: 
$$
\sigma _{\mathrm{k}}\approx 0.69\times 10^{10}\frac{\sigma _{\alpha }}{\sqrt{%
N_{\mathrm{y}}(N_{\mathrm{m}}-1)}}.
$$
where $N_{\mathrm{y}}$ is the number of years for which data is taken. The
total number of measurements is therefore $N_{\mathrm{y}}N_{\mathrm{m}}$.
Indirect constraints currently have a sensitivity no better than $\sigma _{\mathrm{k}%
}=2.3\times 10^{-9}$ \citep{dent}. This would be surpassed by direct
measurements if $\sigma _{\alpha }<2.5\sqrt{N_{\mathrm{y}}(N_{\mathrm{m}}-1)}%
\times 10^{-19}$. For example, if measurements are taken every 20 days (or so)
over a single year ($N_{\mathrm{m}}=18$, $N_{\mathrm{y}}=1$), we need $\sigma _{\alpha }\lesssim 10^{-18}$.  \citet{flamth} have noted that
first excited states in the ${}^{229}\mathrm{Th}$ nucleus is particularly
sensitive to changes in $\alpha $ and $\mu $ with $S_{\alpha }\sim 10^{5}$, allowing for a $\sigma_{\alpha} \lesssim 10^{-20}$.  Such a sensitivity would limit $k_{\alpha}$ at the $10^{-11}$ level -- two orders of magnitude better than the
current WEP violation constraints.

In summary: we have shown how new laboratory constraints on possible time
variation in the fine structure `constant' and the electron-proton mass
ratio can yield more sensitive limits by incorporating the effects of the
seasonal variation of the Sun's gravitational field at the Earth's surface.
This seasonal variation is expected in all theories which require that the
covariant d'Alembertian of any scalar field driving variation of a
`constant' is proportional to the dominant local source of gravitational
potential. The recent experimental results from \citep{Rosenband} and \citep{peik} have reached the
sensitivity of the quasar observations of varying $\alpha $ and $\mu $ made
at high redshift and we have shown may soon provide stronger bounds on
varying constants than conventional ground-based WEP experiments.

\bibliographystyle{aa}

\end{document}